\documentclass[manuscript,screen]{acmart}

\usepackage{multicol}
\usepackage{multirow}
\usepackage{wrapfig}
\usepackage{lscape}
\usepackage{rotating}
\usepackage{amsmath}
\usepackage{bbm}
\usepackage{latexsym}
\usepackage{empheq}

\AtBeginDocument{%
  \providecommand\BibTeX{{%
    \normalfont B\kern-0.5em{\scshape i\kern-0.25em b}\kern-0.8em\TeX}}}

\setcopyright{acmcopyright}
\copyrightyear{2020}
\acmYear{2020}
\acmDOI{10.1145/1122445.1122456}

\begin{document}

\title{Data and knowledge-driven approaches for multilingual training to improve the performance of speech recognition systems of Indian languages}

\author{A. Madhavaraj}
\email{madhavaraja@iisc.ac.in}
\affiliation{%
  \institution{Indian Institute of Science}
  \streetaddress{P.O. Box 1212}
  \city{Bangalore}
  \country{India}
}

\author{Ramakrishnan Angarai Ganesan}
\affiliation{%
  \institution{Indian Institute of Science}
  \streetaddress{P.O. Box 1212}
  \city{Bangalore}
  \state{Karnataka}
  \country{India}}
\email{agr@iisc.ac.in}

\renewcommand{\shortauthors}{Madhavaraj et al.}

\begin{abstract}
We propose data and knowledge-driven approaches for multilingual training of the automated speech recognition (ASR) system for a target language by pooling speech data from multiple source languages. Exploiting the acoustic similarities between Indian languages, we implement two approaches. In phone/senone mapping, deep neural network (DNN) learns non-linear functions to map senones or phones from one language to the others, and the transcriptions of the source languages are modified such that they can be used along with the target language data to train and fine-tune the target language ASR system. In the other approach, we model the acoustic information for all the languages simultaneously by training a multitask DNN (MTDNN) to predict the senones of each language in different output layers. The cross-entropy loss function and the weight update procedure are modified such that only the shared layers and the output layer responsible for predicting the senone classes of a language are updated during training, if the feature vector belongs to that particular language.

In the low-resourced setting (LRS), 40 hours of transcribed speech data each for Tamil, Telugu and Gujarati languages are used for training. The DNN based senone mapping technique gives relative improvements in word error rates (WERs) of 9.66\%, 7.2\% and 15.21\% over the baseline system for Tamil, Gujarati and Telugu languages, respectively. In medium-resourced setting (MRS), 160, 275 and 135 hours of data for Tamil, Kannada and Hindi languages are used, where, the same technique gives better relative improvements of 13.94\%, 10.28\% and 27.24\% for Tamil, Kannada and Hindi, respectively. The MTDNN with senone mapping based training in LRS, gives higher relative WER improvements of 15.0\%, 17.54\% and 16.06\%, respectively for Tamil, Gujarati and Telugu, whereas in MRS, we see improvements of 21.24\% 21.05\% and 30.17\% for Tamil, Kannada and Hindi languages, respectively.

\end{abstract}



\maketitle

\section{Introduction}
\label{sec_1}
To build a good quality speech recognition system, we require a large amount of transcribed speech corpus so that the acoustic model parameters are properly estimated without any over-fitting or under-fitting problems. Although such large scale corpora exist for English and a few other languages, there are no readily available corpora for Indian languages and hence, they are termed as low-resourced languages \cite{madhav_indicon, madhav_interspeech}. Further, collecting such data is a cumbersome and time-consuming task. For low-resourced languages, the traditional way of training the acoustic model using less data results in a high word error rate (WER). If we hypothesize that there exists similarity in phonetic units across languages \cite{vijay2016indicon}, then it is possible to use data from a high-resourced language or multiple low-resourced languages to train the acoustic model of a low-resourced target language and improve its WER performance. In this chapter, we focus exclusively on strategies for training the deep neural network (DNN)-based acoustic model of a target language by exploiting transcribed speech corpora from other source languages.

We review some of the key works in the literature, which employ cross-lingual training. In \cite{lal2013cross}, Lal and King have pooled data from all the source languages and used the perceptive linear prediction (PLP) coefficients and articulatory features and trained a cross-lingual DNN to predict the phone labels. The output from this neural network along with the Mel frequency cepstral coefficient (MFCC) feature vectors in tandem are used as features to train hidden Markov model (HMM) based acoustic models.

In \cite{schultz2001language}, Schultz and Weibel have used data from 5 different languages for 3 categories of multilingual systems namely, ML-sep, ML-mix and ML-tag. ML-sep estimates the acoustic models separately for each of the language, but uses the data pooled from all the languages at the feature extraction stage by computing the feature transformation matrix from the pooled data. ML-mix models the gaussian mixture models (GMMs) by pooling all the data and jointly estimating the GMM weights, means and covariance matrices. ML-tag models the GMM means and covariance matrices from the pooled data but their weights are learnt individually for each language. It can be said that our MT-DNN training strategy explained in section \ref{sec_4} is analogous to the ML-tag modeling, wherein the difference is that the former uses DNN and the latter uses GMMs for modeling the posterior densities.

Various DNN architectures have been proposed to enhance the performance of acoustic models under low-resource scenarios. In \cite{miao2013maxout}, Miao et al. have used deep maxout networks (DMN), where the activations of each hidden layer are grouped and passed through a max-pooling layer before being passed on to the next hidden layer. The advantage of DMNs is that the number of network parameters to be learned is much less than that of the traditional DNNs. Moreover, the maxout activation can approximate any convex function, if the weight vectors are appropriately tuned, thus enabling the network to capture the speech variability in limited data condition.

Mohan and Rose \cite{mohan2015multi} have trained a multitask DNN such that the network parameters are learnt from multiple language data to predict the senone posteriors. After training, the output layers are removed and a new output layer for the target language is created and fine-tuned with the target language data. They have also experimented the effectiveness of low rank approximation of the weight matrices in the DNN and obtained a reduction of 44\% in the number of parameters, and showed a competitive WER performance with respect to the baseline system. The MT-DNN architecture explored in section \ref{sec_4_3} is closely related to the works reported in \cite{mohan2015multi, heigold2013multilingual}.

The motivation to use transcribed speech corpora of other Aryo-Dravidian languages is that there exists similarity in the phonology of Indian languages as reported in \cite{vijay2016indicon}. Further, when data is pooled from closely related languages, better recognition accuracy can be obtained, and has been proven useful for low-resourced languages \cite{datapooling, kvr_is18}. Based on our preliminary study on the phonesets and phonation rules of various Indian languages, we see that about 90\% of the phones are common across Tamil, Kannada, Telugu, Gujarati and Hindi languages. Hence, for improving the performance of the automatic speech recognition (ASR) system for a target language, we can leverage the acoustic information from other languages for better DNN-based acoustic modeling. Towards this purpose, in this work, we have explored two different approaches namely, (i) DPPSM, which leverages information from data of other languages at the acoustic feature-level, and (ii) MT-DNN, which leverages and models the acoustic information at the model-level.

We extend our previous work on multilingual training \cite{madhav_multi} by proposing new phone and senone mapping functions and using this mapping function in multitask DNN training to obtain better ASR performance. We also empirically test the efficacy of the proposed techniques by collecting additional data for other languages like Kannada and Hindi and using them in medium-resourced training setting. This paper is organized as follows: section \ref{sec_1} discusses various methods available in the literature to perform multilingual ASR training. Section \ref{sec_2} presents the details about the datasets used in our experiments, and the procedure to build the baseline DNN based acoustic models. In sections \ref{sec_3} and \ref{sec_4}, we present the proposed techniques of data pooling with phone/senone mapping and multitask DNN training. The WER performances of the proposed techniques in low and medium resourced conditions are given in section \ref{sec_5}. Section \ref{sec_6} concludes the paper.

\section{Dataset used and baseline ASR setup}
\label{sec_2}
\subsection{Dataset used for the study}
\label{sec_2_1}
In this section, we describe the transcribed speech dataset which we have used to build our baseline system as well as for multilingual training. The proposed multilingual training schemes have been tested in low and medium resourced settings. For low-resourced setting, we have used Microsoft's low-resourced Indian language speech recognition corpus for Tamil, Telugu and Gujarati languages \cite{microsoftdata}. In the case of medium resourced setting, we have collected and used our own high quality, transcribed speech datasets for Tamil, Kannada and Hindi languages. The reason for having different sets of languages in low and medium resourced setting is to test the robustness of the proposed techniques, at the same time proving the hypothesis that similarity exists in the phonology of Indian languages as reported in \cite{vijay2016indicon}. The details of the duration, number of utterances, and the number of speakers for low and medium resourced datasets are listed in Tables \ref{tab_1} and \ref{tab_2}, respectively.

\begin{table*}[htbp]
  \caption {Details of the dataset (provided by Microsoft) used for multilingual DNN training experiments under low-resource setting specifying the number of hours and utterances.}
  \centering
  \resizebox{\textwidth}{0.08\textheight}{ \begin{tabular}{| c | c c c | c c c | c c c | }
  \hline
  \multirow{3}{*}{\textbf{Specs.}} & \multicolumn{3}{c|}{\textbf{Tamil}} & \multicolumn{3}{c|}{\textbf{Telugu}} & \multicolumn{3}{c|}{\textbf{Gujarati}} \\[2pt]
  \cline{2-10}
  & Train & Dev. & Test & Train & Dev. & Test & Train & Dev. & Test \\[4pt]
  \hline
  \hline
  Hours & 40 & 4.3 & 5 & 40 & 4.2 & 5 & 40 & 5 & 5 \\[4pt]
  \hline
  Utterances & 39131 & 2609 & 3081 & 44882 & 2549 & 3040 & 22807 & 3075 & 3419 \\[4pt]
  \hline
  \end{tabular}%
  }
  \label{tab_1}
\end{table*}

\begin{table*}[htbp]
  \caption {Details of the dataset used for multilingual DNN training experiments under medium-resource setting specifying the number of hours, speakers and utterances.}
  \centering
  \resizebox{\textwidth}{0.09\textheight}{ \begin{tabular}{| c | c c c | c c c | c c c | }
  \hline
  \multirow{3}{*}{\textbf{Specs.}} & \multicolumn{3}{c|}{\textbf{Tamil}} & \multicolumn{3}{c|}{\textbf{Kannada}} & \multicolumn{3}{c|}{\textbf{Hindi}} \\[2pt]
  \cline{2-10}
  & Train & Dev. & Test & Train & Dev. & Test & Train & Dev. & Test \\[4pt]
  \hline
  \hline
  Hours & 160 & 7 & 50 & 275 & 36 & 37 & 135 & 26 & 31 \\[4pt]
    \hline
  Speakers & 440 & 20 & 140 & 700 & 90 & 92 & 205 & 47 & 54 \\[4pt]
  \hline
  Utterances & 52800 & 2400 & 16800 & 83652 & 10808 & 11084 & 87425 & 16860 & 20607 \\[4pt]
  \hline
  \end{tabular}%
  }
  \label{tab_2}
\end{table*}

\subsection{Baseline ASR setup}
\label{sec_2_2}
We have built DNN-based ASR systems separately for each language for both low and medium resourced conditions and used them as baseline systems for all our experiments. We have used our own grapheme-to-phoneme mapping tool to convert the Unicode text transcriptions to the corresponding phone sequences. First we train a simple left-to-right, 3-state monophone hidden Markov model for each phone and model each HMM state density as a Gaussian mixture model. The model parameters are learnt using expectation maximization (EM) procedure for 40 iterations. Using the monophone alignments, we next build context-dependent triphone models by taking each phone's left and right contexts into account. Then, using the triphone alignments, we build linear discriminant analysis and maximum likelihood linear transformation (LDA-MLLT) based HMM models. Using the phone-level alignments of LDA-MLLT models, we perform speaker adaptive training (SAT) to refine the alignments further. These SAT alignments are then used to train a DNN model containing 7 layers with each layer containing 1024 nodes for 16 epochs. During training, the initial learning rate is set as 0.08, which is halved after every epoch. This DNN acoustic model is combined with 3-gram language model estimated from the text corpus to obtain the final baseline ASR system. More details about our baseline training setup can be seen in \cite{madhav_indicon}.

\section{Data pooling with phone/senone mapping methods}
\label{sec_3}
In most of the multilingual training techniques in the literature, phone mapping across languages is achieved through the use of an universal phoneset \cite{ipa, cmu_frontend}. In this technique, we map the phones from the source languages to the target language using a DNN trained on the target language data. The main advantage of the proposed mapping technique is that it uses the acoustic information (i.e., the feature vectors) from the source languages' data to automatically create the phone map in a data-driven fashion rather than using a predefined manually created phone map table.

The phonetic transcription of each source language data is then modified using the phone map table to suit the training requirements of the target language ASR \cite{agr2007g2p, agr2015translit}. We then train a DNN acoustic model by pooling together data from all the languages. The DNN is then fine-tuned for a few epochs using only the target language data. The detailed procedure to perform multilingual training based on data pooling is given in the following subsections.

\subsection{Initial DNN training for phone mapping}
\label{sec_3_1}
The first step is to train a DNN acoustic model from the target language data, and then use it to map the phones from any source language to the target language phones. We follow the standard procedure explained in section \ref{sec_2_2} to create the DNN acoustic model for the target language. This DNN can now predict the target language senones for the given input acoustic feature vectors.

\subsection{Generating alignments for the source language data}
\label{sec_3_2}
The next step is to create the senone alignments for the source language data, for which we train an independent SAT model for each of the source languages. We then align the source language data to generate the senone alignments independently for each of the source languages using the corresponding SAT model. Henceforth, we refer to them as source alignments.

\subsection{Phone mapping from the source to the target language}
\label{sec_3_3}
Since the acoustic model for each language is independently built using its own respective phoneset and senones, there is a need to map these entities from one language to the other in order to pool the data together and train the ASR system for any target language. For this purpose, we have employed two different types of mapping, namely senone-level and phone-level, where the former is a finer mapping while the latter is a coarser mapping of the phonetic units.

\begin{figure*}[!ht]
\includegraphics[width=0.92\textwidth,height=0.68\textheight]{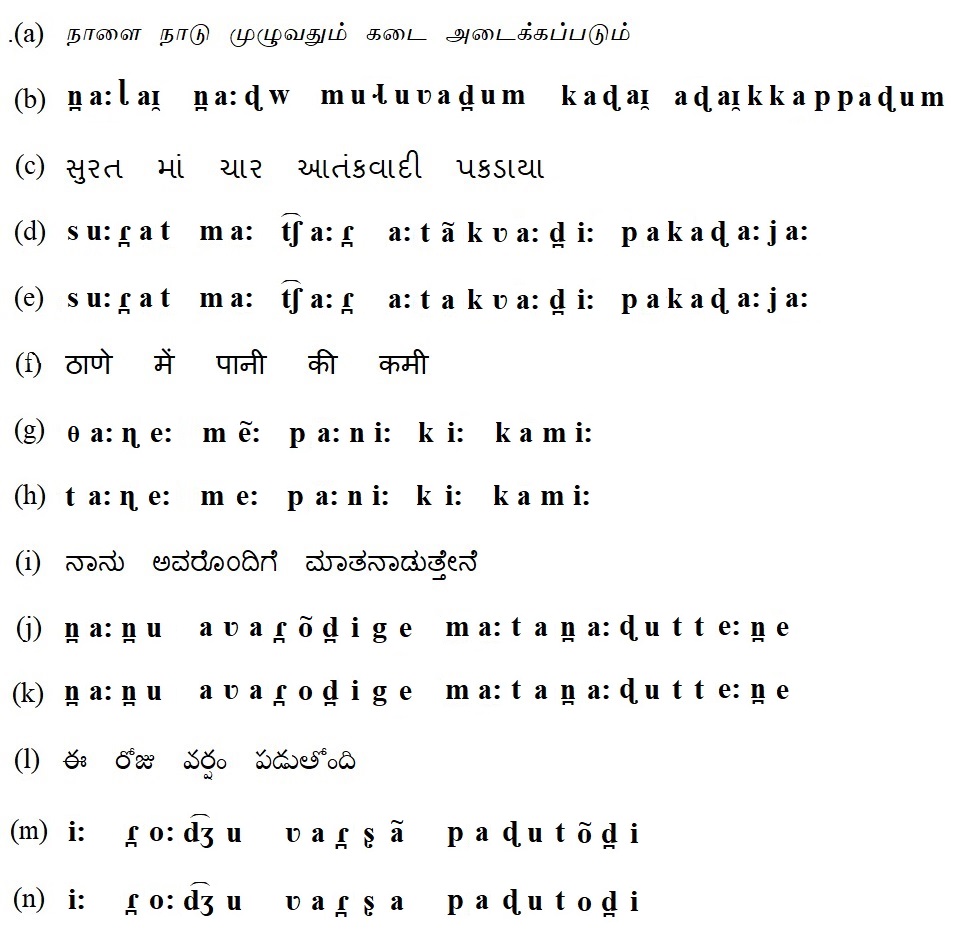}
\centering
\caption{Mapping of phones of different source languages to the target language's (Tamil) phones using manual mapping method : (a-b) A sample Tamil sentence and its phone sequence, (c-e) A sample Gujarati sentence, its corresponding phone sequence and phone sequence mapped to Tamil phone sequence, (f-h) A sample Hindi sentence, its phone sequence and mapped phone sequence, (i-k) A sample Kannada sentence, its phone sequence and mapped phone sequence, (l-n) A sample Telugu sentence, its phone sequence and mapped phone sequence.}
	\label{fig_1}
\end{figure*}

\vspace{1cm}
\noindent
\textbf{Senone-level mapping:} The senones ($S_{s_{l}}$) of the source language $s_{l}$ are mapped to the senones ($S_T$) of the target language $T$ by passing the input acoustic features $X_{s_l}$ of the source language to the DNN model of the target language and obtaining the predicted label $f(X_{s_l})$. Let the actual senone label for the feature vector $X_{s_l}$ be denoted as $O(X_{s_l})$. We calculate the conditional density $p(y_T|y_{s_{l}})$ using the following equation,

\begin{align}
    \label{eqn_1}
    p_s(y_T | y_{s_l}) &= \frac{\sum_{X_{s_l}} \mathbbm{1} \left[ f(X_{s_l}) = y_T \; : \; O(X_{s_l}) = y_{s_l} \right ] } {\sum_{X_{s_l}} \mathbbm{1} \left[ O(X_{s_l}) = y_{s_l} \right ] } \\[10pt]
    \intertext{\indent where $y_T$ can take any value from the senone-set $S_{T}$, and $y_{s_l}$ can take any value from the senone-set $S_{s_l}$ and $\mathbbm{1}\left[ \cdot \right]$ is the indicator function. Now, the final senone mapping function $\psi_s(y_{s_l})$ is calculated by taking $argmax$ over $y_T$ on this conditional density.}
    \label{eqn_2}
    \psi_s(y_{s_l}) &= y_{T}^{*} = argmax_{y_T} \; p_s(y_T | y_{s_l})
\end{align}

This mapping function maps the senone $y_{s_l}$ of the source language to that particular senone $y_{T}^{*}$ of the target language, which is predicted maximum number of times for the feature vectors belonging to the senone $y_{s_l}$.

\vspace{1cm}
\noindent
\textbf{Phone-level mapping:} In this mapping scheme, we directly map the phones of any source language to the phones of the target language through the senone alignments using the senone-to-phone mapping functions $g_{s_l}(\cdot)$ and $g_{T}(\cdot)$ of the source and target languages, respectively. These senone-to-phone mapping functions can directly be obtained from the decision tree built while clustering the HMM states. Let $\Phi_{s_l}$ and $\Phi_T$ be the phonesets of the source language $s_l$ and target language $T$, respectively. Then, the phone mapping function $\psi_{pm}(\cdot)$ can be derived as,

\begin{align}
    \label{eqn_3}
    p_{pm}(z_T | z_{s_l}) &= \frac{\sum_{X_{s_l}} \mathbbm{1} \left[ g_{T}(f(X_{s_l})) = z_T : g_{s_l}(O(X_{s_l})) = z_{s_l} \right ] } {\sum_{X_{s_l}} \mathbbm{1} \left[ g_{s_l}(O(X_{s_l})) = z_{s_l} \right ] } \\[10pt]
    \intertext{\indent where $z_T$ can take any value from the phoneset $\Phi_{T}$ of the target language, and $z_{s_l}$ can take any value from the phone-set $\Phi_{s_l}$ of the source language $s_l$. Now, the final phone mapping $\psi_{pm}(y_{s_l})$ is calculated by taking $argmax$ over $z_T$ on this conditional density:}
    \label{eqn_4}
    \psi_{pm}(z_{s_l}) &= z_{T}^{*} = argmax_{z_T} \; p_{pm}(z_T | z_{s_l})
\end{align}

Thus using $\psi_{s}(\cdot)$ and $\psi_{pm}(\cdot)$, we can convert the transcription from any source language to the target language at senone-level and phone-level, respectively.

Additionally, using prior knowledge of the phonation rules of the languages considered in our experiment, we have also created a manual phone mapping table, which converts the phone sequence of any language to the phone sequence of any other language. Figure \ref{fig_1} shows an example Tamil sentence and its phone sequence, and one sample sentence each from Gujarati, Hindi, Kannada and Telugu and their corresponding phone sequences and their mapped versions in the target language, namely Tamil.

\subsection{Data pooling and training}
\label{sec_3_4}
Once the transcriptions of all the source language data are converted using $\psi_{s}(\cdot)$, $\psi_{pm}(\cdot)$ or the manual mapping table to the format required to train the target language ASR, we pool them along with the target language data and train a new DNN for 16 epochs. Since more training utterances are available due to the data pooling, the modeling of acoustic information is expected to be better than that in the baseline system, where only the target language data is used for training.

\subsection{DNN fine-tuning for the target language}
\label{sec_3_5}
The DNN model obtained in the previous stage is fine-tuned with only the target language data for an additional five epochs with a learning rate of 0.0008. This fine-tuned DNN is now used as the final acoustic model for decoding the test utterances of the target language.

\section{Multitask DNN training methods}
\label{sec_4}
\begin{figure*}[!ht]
\includegraphics[width=0.9\textwidth,height=0.76\textheight]{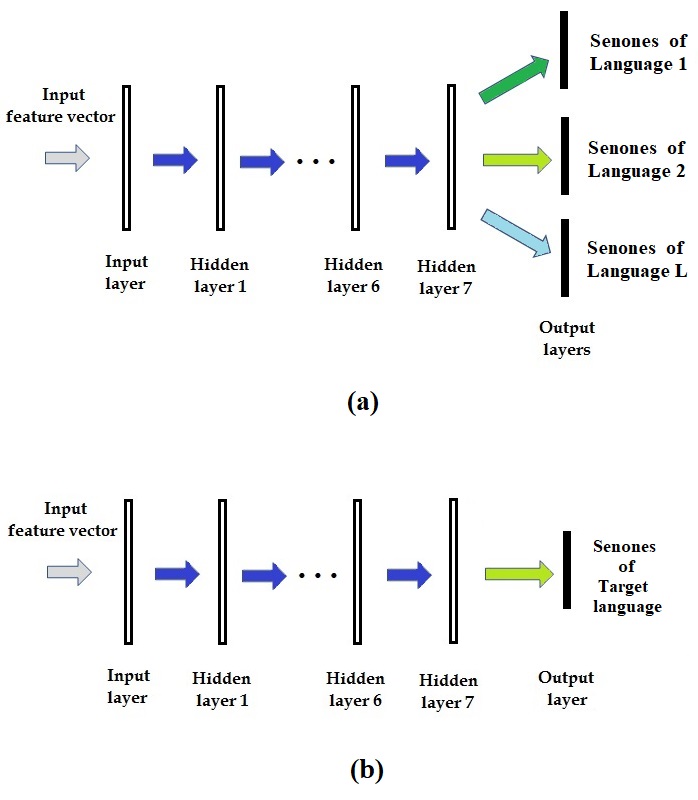}
\centering
\caption{(a) Multitask DNN architecture to predict senones for the three languages for any given input vector. The shared hidden layers learn the feature representations common to all the languages and the three output layers perform the classification for the individual languages. (b) The feed-forward DNN pruned from the trained multitask DNN, which needs to be fined-tuned to predict the senones of the target language.}
	\label{fig_2}
\end{figure*}

In this method, we have used a multitask deep neural network (MT-DNN), which is a cascade of many hidden layers, and contains as many output layers as the number of languages used for training. The architecture of MT-DNN is shown in Fig. \ref{fig_2}a, which is similar to the one in \cite{huang2013shldnn}. This network is trained by pooling the data of all the languages in a specific manner. The cross-entropy loss function is modified in two different ways and used for training. Finally, we prune the undesired output layers from the MT-DNN and construct a simple feed-forward DNN and fine-tune it only with the target language data to get the final DNN acoustic model (AM) and then use it for decoding. The steps are explained below in detail.

\subsection{Generating alignments for the source and target languages}
\label{sec_4_1}
We have built the SAT models for the source and the target languages independently following the procedure same as in section \ref{sec_3_1}, and using these models, senone alignments are generated. Since the senones differ in number and correspondence across languages, we learn the correspondence implicitly at the model-level through the MT-DNN by training it in a specific manner as illustrated below.

\subsection{Training the multitask DNN}
\label{sec_4_2}
We have experimented with two different ways of providing the features and labels to train the MT-DNN. The first method provides target labels only to the output layer corresponding to the language to which the feature vector belongs. The rest of the output layers are not made to predict anything and hence no loss is back-propagated from them. The second method provides target labels to all the output layers using the senone-map function $\phi_{S_i}(\cdot)$ obtained in section \ref{sec_3_3}. These two methods are formulated as follows.

\vspace{0.5cm}
\noindent
\textbf{Target labels for only one output layer:} Let $x$ be the feature vector belonging to the language $l$ and $s$ be its corresponding senone target label. Each training example to the MT-DNN should be of the form $\{ x,[{y_1},{y_2},...,{y_L}]\}$, where $y_l$ is the desired target in one-hot vector encoding format for the $l^{th}$ output layer and $L$ is the total number of languages. The $i^{th}$ entry of the target label vector $y_l^i$ takes the value 1 only when the feature vector belongs to the particular language $l$ and senone label $i$; else it is 0. Thus, it is defined as,

\begin{equation}
    \begin{array}{*{20}{c}}
{y_l^i = \mathbbm{1} \left [ x \in l \right ] \mathbbm{1} \left [ s = i \right ] }&{ \; \forall 1 \le l \le L}
    \end{array}
    \label{eqn_5}
\end{equation}

We can see that a feature vector belonging to a particular language is assigned zero as the desired target for senones of all the other languages \cite{heigold2013multilingual}. However, in the context of MT-DNN, for a given feature vector, it is inappropriate to make the MT-DNN to forcefully predict zero as the senone-posterior for the other languages. Hence, we modify the loss function in such a way that only those layers are updated that are responsible for predicting the senones for the language to which the feature vector belongs. This modified loss function at the $l^{th}$ output layer ${\hat{\mathcal{L}^l}}$ is given by,

\begin{equation}
    {\hat {\mathcal{L}^l}}({z_l},{y_l}) = \mathcal{L}({z_l},{y_l})\mathbbm{1} \left [ x \in l \right ]
    \label{eqn_6}
\end{equation}

where $z_l$ is the actual predicted vector, $y_l$ is the desired target vector at the $l^{th}$ layer and $\mathcal{L}(\cdot)$ is the standard cross entropy loss function.

\vspace{0.5cm}
\noindent
\textbf{Target labels for all the output layers:} First we create the set of senone map functions for all possible pairs of languages:

\begin{equation}
    \left \{ \phi_S^l \left ( y_{s_m} \right) \;\; : \;\; 1\leq l,m \leq L \right \}
    \label{eqn_4_7}
\end{equation}

Using these mapping functions, we can now map senones from any language $l$ to any other language $m$. If $l = m$, then the mapping is an identity function. For a given feature vector $x$ and its corresponding target senone label $s$, the input examples to be fed to the MT-DNN are represented as $\{ x,[{y_1},{y_2},...,{y_L}]\}$ where,

\begin{equation}
    \begin{array}{*{20}{c}}
{y_l^i = \mathbbm{1}( \phi_S^l (s) = i)}&{ \; \forall 1 \le l \le L}
    \end{array}
    \label{eqn_8}
\end{equation}

Now, we have non-zero targets at every output layer for any given feature vector. This way of setting the target labels at all the output layers can be thought of as a hybrid combination of senone mapping and vanilla MT-DNN approach. Standard cross entropy loss is used as the loss function for this setup.

Next, we train the MT-DNN with 7 hidden layers using this modified loss function with the feature vectors and targets (either by using equation \ref{eqn_5} or \ref{eqn_8}) by pooling all the data together. The MT-DNN is trained for 16 epochs using Keras \cite{keras} library. The initial learning rate is fixed at 0.008 and then halved after every epoch.

The DNN architecture is trained in this fashion to ensure that all the hidden layers learn feature representations that are common to all the languages, while increasing the discriminability of every output layer.

\subsection{Pruning and fine-tuning of multitask DNN for the target language}
\label{sec_4_3}
Next, we take the trained MT-DNN network and prune it so that we keep only the hidden layers and the output layer that predicts the senones of the target language, as shown in figure \ref{fig_2}b. This feed-forward DNN is then fine-tuned for an additional 5 epochs with the data from only the desired target language with a learning rate of 0.0008. This fine-tuned network is now used as the final acoustic model for decoding the test utterances.

\section{Experimental results}
\label{sec_5}
We have created a multilingual training setup and trained ASR systems for the target languages using different phone/senone mapping schemes explained in section \ref{sec_3}. These systems have been trained and tested in low and medium-resource settings and the WERs are compared with those of the baseline systems and tabulated in Table \ref{tab_3}. The Table shows that the DNN-based senone mapping gives the best WERs of 29.94\% and 11.67\% on the test sets compared to the baseline WERs of 33.14\% and 13.56\% in low and medium-resource conditions, respectively. We obtain relative improvements of 6.78\% and 9.03\% over the baselines on the development datasets as well.

The DNN based phone mapping performs marginally better than the manual phone mapping; and both these schemes perform better than the baseline system. Table \ref{tab_3} shows that the senone mapping method performs far better than both the phone mapping methods. With these results, we can infer that finer mapping of transcription from source languages to the target language helps in better acoustic modeling in a multilingual training scenario.

\begin{table*}[htbp]
  \caption {Comparison of WERs (in \%) of the Tamil ASR for the baseline systems and data pooling with phone/senone mapping techniques under low (Tamil, Telugu, Gujarati data) and medium-resource (Tamil, Kannada, Hindi data) conditions. Relative improvements in the WER with respect to the baseline are given in parentheses for each case.}
\centering 
\resizebox{\textwidth}{0.16\textheight}{
\begin{tabular}{| c | c | c | c | }
 \hline
  \textbf{Resource} & \textbf{Method} & \textbf{Dev. set WER} & \textbf{Test set WER} \\[2pt]
  \hline
  \hline
  \multirow{4}{*}{\textbf{Low}} & Baseline Tamil ASR & 32.87 (NA) & 33.14 (NA) \\[4pt]
  \cline{2-4}
   & Manual phone mapping & 32.33 (1.64) & 32.50 (1.93) \\[4pt]
   \cline{2-4}
   & DNN-based phone mapping & 32.31 (1.70) & 32.16 (2.96) \\[4pt]
   \cline{2-4}
   & DNN-based senone mapping & 30.64 (6.78) & 29.94 (9.66) \\[4pt]
  \hline
  \hline
  \multirow{4}{*}{\textbf{Medium}} & Baseline Tamil ASR & 16.28 (NA) & 13.56 (NA) \\[4pt]
  \cline{2-4}
   & Manual phone mapping & 15.14 (7.0) & 12.94 (4.57) \\[4pt]
   \cline{2-4}
   & DNN based phone mapping & 15.07 (7.43) & 12.92 (4.72) \\[4pt]
   \cline{2-4}
   & DNN based senone mapping & 14.81 (9.03) & 11.67 (13.94) \\[4pt]
  \hline
\end{tabular}%
  }
  \label{tab_3}
\end{table*}

The MT-DNN based multilingual training, which uses the advantages of changes to both the DNN architecture and loss function, aids in better prediction of senones of the target language, thereby improving the performance as seen in Table \ref{tab_4}. The simple MT-DNN gives WERs of 32.07\% and and 11.24\%, whereas the baseline WERs are 33.14\% and 13.56\% in low and medium-resource conditions, respectively. Incorporating the advantages of senone mapping scheme to MT-DNN training gives the best WERs of 28.17\% and 10.68\% in low and medium-resource conditions, respectively, on the test datasets. We see similar WER improvements on the development set too.

\begin{table*}[htbp]
  \caption {Comparison of Tamil ASR WERs (in \%) for the baseline systems and multitask DNN (MT-DNN) training techniques under low (Tamil, Telugu, Gujarati data) and medium-resource (Tamil, Kannada, Hindi data) conditions. Relative improvements in WER with respect to the baseline are given in parentheses for each case.}
\centering 
\resizebox{\textwidth}{0.13\textheight}{
\begin{tabular}{| c | c | c | c | }
 \hline
  \textbf{Resource} & \textbf{Method} & \textbf{Dev. set WER} & \textbf{Test set WER} \\[2pt]
  \hline
  \hline
  \multirow{4}{*}{\textbf{Low}} & Tamil Baseline ASR & 32.87 (NA) & 33.14 (NA) \\[4pt]
  \cline{2-4}
   & Simple MT-DNN & 31.35 (4.62) & 32.07 (3.22) \\[4pt]
   \cline{2-4}
   & MT-DNN with senone mapping & 29.03 (11.68) & 28.17 (15.0) \\[4pt]
  \hline
  \hline
  \multirow{4}{*}{\textbf{Medium}} & Tamil Baseline ASR & 16.28 (NA) & 13.56 (NA) \\[4pt]
  \cline{2-4}
   & Simple MT-DNN & 15.01 (7.8) & 11.24 (17.11) \\[4pt]
   \cline{2-4}
   & MT-DNN with senone mapping & 13.74 (15.6) & 10.68 (21.24) \\[4pt]
  \hline
\end{tabular}%
  }
  \label{tab_4}
\end{table*}

\begin{figure*}[!ht]
\includegraphics[width=\textwidth,height=0.4\textheight]{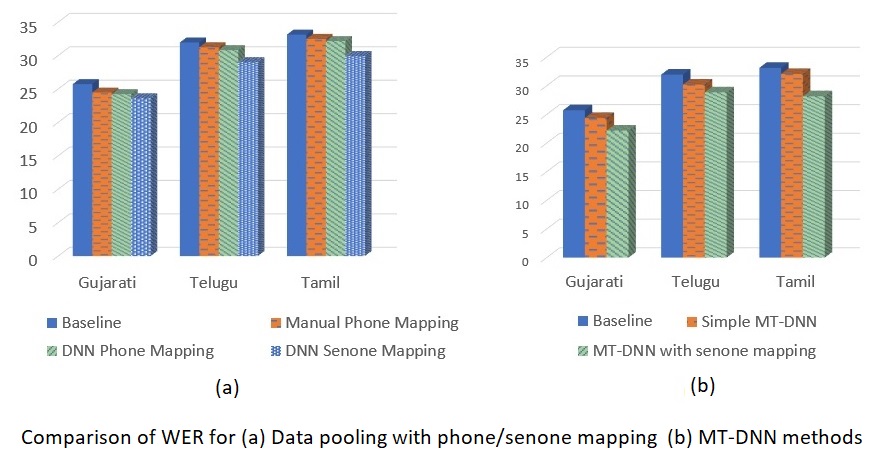}
\centering
\caption{Performance comparison of (a) data pooling with phone/senone mapping and (b) multitask DNN methods under low-resource condition for Gujarati, Telugu and Tamil languages.}
	\label{fig_3}
\end{figure*}

\begin{figure*}[!ht]
\includegraphics[width=\textwidth,height=0.4\textheight]{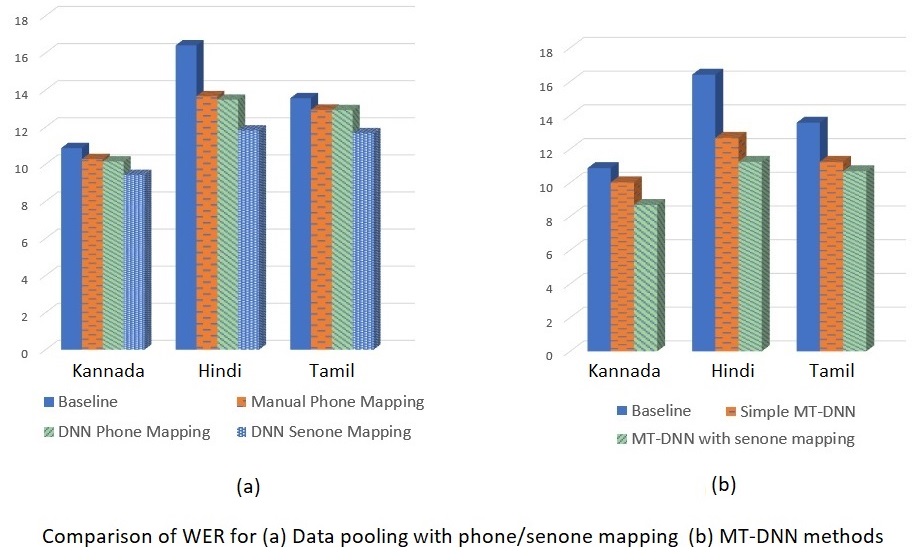}
\centering
\caption{Performance comparison of (a) data pooling with phone/senone mapping and (b) multitask DNN methods under medium-resource condition for Kannada, Hindi and Tamil languages.}
	\label{fig_4}
\end{figure*}

We have also tested the proposed multilingual training methods by training ASR systems for all the combinations of source and target languages. Figures \ref{fig_3} and \ref{fig_4} show the performances of all the variants of DP-PSM and MT-DNN training methods in low and medium-resource conditions. We see that for all the languages, the proposed techniques perform better than their respective baseline systems. Thus, we say that the similarity across Indian languages can be exploited using DP-PSM and MT-DNN methods to build speech recognition system for any target Indian language, in a multilingual scenario with relatively limited data.

\section{Conclusion}
\label{sec_6}
Thus, we have proposed two different approaches, namely data pooling with phone/senone mapping and multitask DNN to develop ASR systems of a traget language by leveraging acoustic information from transcribed speech corpora of other source languages. We have used transcribed speech corpus of Tamil, Telugu, Kannada, Hindi and Gujarati languages and both these approaches have been empirically tested and shown to perform better than the baseline systems under low and medium-resource constraints.

The DP-PSM approach pools the data together by mapping the phones/senones from the source languages to the target language. The mapping functions are derived either based on known phonation rules of the languages (knowledge-driven) or by a data-driven manner using a pre-trained target language DNN. The pooled data is now used to train a DNN acoustic model and then fine-tuned with the target language data. For the target Tamil language ASR, the senone mapping method gives WERs of 29.94\% and 11.67\%, compared to the baseline WERs of 33.14\% and 13.56\% in low and medium-resource settings, respectively. This method outperforms both manual and data-driven phone-based mapping methods. Similar trend is seen when we take any other combination of source and target languages, in low or medium-resource settings.

Alternatively, we have seen that the MT-DNN approach based multilingual training can be coupled with senone mapping to set targets for all the output layers. Accordingly, the cross entropy loss function is modified to properly train the MT-DNN. Such a training strategy employs the benefits of both DP-PSM and MT-DNN and gives relative WER improvements of 15.0\% and 21.24\% over the respective low and medium-resourced baseline Tamil ASR systems. Similar improvements are observed for other combinations of source and target languages as well.

\bibliographystyle{ACM-Reference-Format}
\bibliography{sample-manuscript}


\begin{thebibliography}{18}


\ifx \showCODEN    \undefined \def \showCODEN     #1{\unskip}     \fi
\ifx \showDOI      \undefined \def \showDOI       #1{#1}\fi
\ifx \showISBNx    \undefined \def \showISBNx     #1{\unskip}     \fi
\ifx \showISBNxiii \undefined \def \showISBNxiii  #1{\unskip}     \fi
\ifx \showISSN     \undefined \def \showISSN      #1{\unskip}     \fi
\ifx \showLCCN     \undefined \def \showLCCN      #1{\unskip}     \fi
\ifx \shownote     \undefined \def \shownote      #1{#1}          \fi
\ifx \showarticletitle \undefined \def \showarticletitle #1{#1}   \fi
\ifx \showURL      \undefined \def \showURL       {\relax}        \fi
\providecommand\bibfield[2]{#2}
\providecommand\bibinfo[2]{#2}
\providecommand\natexlab[1]{#1}
\providecommand\showeprint[2][]{arXiv:#2}

\bibitem[\protect\citeauthoryear{??}{mic}{2018}]%
        {microsoftdata}
 \bibinfo{year}{2018}\natexlab{}.
\newblock \bibinfo{booktitle}{\emph{Data provided by {SpeechOcean.com} and
  {M}icrosoft}}.
\newblock \bibinfo{publisher}{Microsoft}.
\newblock


\bibitem[\protect\citeauthoryear{Chellapriyadharshini, Toffy, Srinivasa K~M,
  and Ramasubramanian}{Chellapriyadharshini et~al\mbox{.}}{2018}]%
        {kvr_is18}
\bibfield{author}{\bibinfo{person}{Maharajan Chellapriyadharshini},
  \bibinfo{person}{Anoop Toffy}, \bibinfo{person}{Raghavan Srinivasa K~M},
  {and} \bibinfo{person}{V Ramasubramanian}.} \bibinfo{year}{2018}\natexlab{}.
\newblock \showarticletitle{Semi-supervised and active-learning scenarios:
  Efficient acoustic model refinement for low resource {Indian} language}. In
  \bibinfo{booktitle}{\emph{19th Annual Conference of the International Speech
  Communication Association (Interspeech 2018)}}.
\newblock


\bibitem[\protect\citeauthoryear{Chollet et~al\mbox{.}}{Chollet
  et~al\mbox{.}}{2015}]%
        {keras}
\bibfield{author}{\bibinfo{person}{Fran{\c{c}}ois Chollet} {et~al\mbox{.}}}
  \bibinfo{year}{2015}\natexlab{}.
\newblock \showarticletitle{keras. GitHub repository}.
\newblock \bibinfo{journal}{\emph{https://github. com/fchollet/keras>. Accessed
  on}}  \bibinfo{volume}{25} (\bibinfo{year}{2015}), \bibinfo{pages}{2017}.
\newblock


\bibitem[\protect\citeauthoryear{Heigold, Vanhoucke, Senior, Nguyen, Ranzato,
  Devin, and Dean}{Heigold et~al\mbox{.}}{2013}]%
        {heigold2013multilingual}
\bibfield{author}{\bibinfo{person}{Georg Heigold}, \bibinfo{person}{Vincent
  Vanhoucke}, \bibinfo{person}{Alan Senior}, \bibinfo{person}{Patrick Nguyen},
  \bibinfo{person}{Marc’Aurelio Ranzato}, \bibinfo{person}{Matthieu Devin},
  {and} \bibinfo{person}{Jeffrey Dean}.} \bibinfo{year}{2013}\natexlab{}.
\newblock \showarticletitle{Multilingual acoustic models using distributed deep
  neural networks}. In \bibinfo{booktitle}{\emph{Acoustics, Speech and Signal
  Processing (ICASSP), IEEE International Conf. on}}.
  \bibinfo{pages}{8619--8623}.
\newblock


\bibitem[\protect\citeauthoryear{Hieronymus}{Hieronymus}{1994}]%
        {ipa}
\bibfield{author}{\bibinfo{person}{J~L Hieronymus}.}
  \bibinfo{year}{1994}\natexlab{}.
\newblock \bibinfo{title}{{ASCII} Phonetic Symbols for the World's Languages:
  Worldbet}.
\newblock
\newblock


\bibitem[\protect\citeauthoryear{Huang, Li, Yu, Deng, and Gong}{Huang
  et~al\mbox{.}}{2013}]%
        {huang2013shldnn}
\bibfield{author}{\bibinfo{person}{Jui-Ting Huang}, \bibinfo{person}{Jinyu Li},
  \bibinfo{person}{Dong Yu}, \bibinfo{person}{Li Deng}, {and}
  \bibinfo{person}{Yifan Gong}.} \bibinfo{year}{2013}\natexlab{}.
\newblock \showarticletitle{Cross-language knowledge transfer using
  multilingual deep neural network with shared hidden layers}. In
  \bibinfo{booktitle}{\emph{Acoustics, Speech and Signal Processing (ICASSP),
  IEEE International Conf. on}}. \bibinfo{pages}{7304--7308}.
\newblock


\bibitem[\protect\citeauthoryear{Lal and King}{Lal and King}{2013}]%
        {lal2013cross}
\bibfield{author}{\bibinfo{person}{Partha Lal} {and} \bibinfo{person}{Simon
  King}.} \bibinfo{year}{2013}\natexlab{}.
\newblock \showarticletitle{Cross-lingual automatic speech recognition using
  tandem features}.
\newblock \bibinfo{journal}{\emph{IEEE Trans. Audio, Speech, and Language
  Processing}} \bibinfo{volume}{21}, \bibinfo{number}{12}
  (\bibinfo{year}{2013}), \bibinfo{pages}{2506--2515}.
\newblock


\bibitem[\protect\citeauthoryear{Madhavaraj and Ramakrishnan}{Madhavaraj and
  Ramakrishnan}{2017}]%
        {madhav_indicon}
\bibfield{author}{\bibinfo{person}{A Madhavaraj} {and} \bibinfo{person}{A~G
  Ramakrishnan}.} \bibinfo{year}{2017}\natexlab{}.
\newblock \showarticletitle{Design and development of a large vocabulary,
  continuous speech recognition system for {Tamil}}. In
  \bibinfo{booktitle}{\emph{2017 14th IEEE India Council International
  Conference (INDICON)}}. IEEE, \bibinfo{pages}{1--5}.
\newblock


\bibitem[\protect\citeauthoryear{{Madhavaraj} and {Ramakrishnan}}{{Madhavaraj}
  and {Ramakrishnan}}{2019}]%
        {madhav_multi}
\bibfield{author}{\bibinfo{person}{A. {Madhavaraj}} {and}
  \bibinfo{person}{A.~G. {Ramakrishnan}}.} \bibinfo{year}{2019}\natexlab{}.
\newblock \showarticletitle{Data-pooling and multi-task learning for enhanced
  performance of speech recognition systems in multiple low resourced
  languages}. In \bibinfo{booktitle}{\emph{2019 National Conference on
  Communications (NCC)}}. \bibinfo{pages}{1--5}.
\newblock


\bibitem[\protect\citeauthoryear{Madhavaraj, Shiva~Kumar, and
  Ramakrishnan}{Madhavaraj et~al\mbox{.}}{2018}]%
        {madhav_interspeech}
\bibfield{author}{\bibinfo{person}{A Madhavaraj}, \bibinfo{person}{H~R
  Shiva~Kumar}, {and} \bibinfo{person}{A~G Ramakrishnan}.}
  \bibinfo{year}{2018}\natexlab{}.
\newblock \showarticletitle{Online speech translation system for {Tamil}}. In
  \bibinfo{booktitle}{\emph{19th Annual Conference of the International Speech
  Communication Association (INTERSPEECH 2018)}}.
\newblock


\bibitem[\protect\citeauthoryear{Miao, Metze, and Rawat}{Miao
  et~al\mbox{.}}{2013}]%
        {miao2013maxout}
\bibfield{author}{\bibinfo{person}{Y. Miao}, \bibinfo{person}{F. Metze}, {and}
  \bibinfo{person}{S. Rawat}.} \bibinfo{year}{2013}\natexlab{}.
\newblock \showarticletitle{Deep maxout networks for low-resource speech
  recognition}. In \bibinfo{booktitle}{\emph{IEEE Workshop on Automatic Speech
  Recognition and Understanding}}. \bibinfo{pages}{398--403}.
\newblock


\bibitem[\protect\citeauthoryear{Mohan and Rose}{Mohan and Rose}{2015}]%
        {mohan2015multi}
\bibfield{author}{\bibinfo{person}{Aanchan Mohan} {and}
  \bibinfo{person}{Richard Rose}.} \bibinfo{year}{2015}\natexlab{}.
\newblock \showarticletitle{Multi-lingual speech recognition with low-rank
  multi-task deep neural networks}. In \bibinfo{booktitle}{\emph{Acoustics,
  Speech and Signal Processing (ICASSP), IEEE International Conf. on}}.
  \bibinfo{pages}{4994--4998}.
\newblock


\bibitem[\protect\citeauthoryear{Parlikar, Sitaram, Wilkinson, and
  Black}{Parlikar et~al\mbox{.}}{2016}]%
        {cmu_frontend}
\bibfield{author}{\bibinfo{person}{Alok Parlikar}, \bibinfo{person}{Sunayana
  Sitaram}, \bibinfo{person}{Andrew Wilkinson}, {and} \bibinfo{person}{Alan~W.
  Black}.} \bibinfo{year}{2016}\natexlab{}.
\newblock \showarticletitle{The {Festvox} {Indic} Frontend for
  {Grapheme-to-Phoneme} Conversion}. In \bibinfo{booktitle}{\emph{Proceedings
  of 10th Language Resources and Evaluation Conference}}.
\newblock


\bibitem[\protect\citeauthoryear{Ramakrishnan and Laxmi~Narayana}{Ramakrishnan
  and Laxmi~Narayana}{2007}]%
        {agr2007g2p}
\bibfield{author}{\bibinfo{person}{A~G Ramakrishnan} {and} \bibinfo{person}{M
  Laxmi~Narayana}.} \bibinfo{year}{2007}\natexlab{}.
\newblock \showarticletitle{Grapheme to phoneme conversion for {Tamil} speech
  synthesis}. In \bibinfo{booktitle}{\emph{Proc. Workshop in Image and Signal
  Processing (WISP-2007), IIT Guwahati}}. \bibinfo{pages}{96--99}.
\newblock


\bibitem[\protect\citeauthoryear{Ramakrishnan, Sequiera, Rao, and
  Shiva~Kumar}{Ramakrishnan et~al\mbox{.}}{2015}]%
        {agr2015translit}
\bibfield{author}{\bibinfo{person}{A~G Ramakrishnan},
  \bibinfo{person}{Royal~Denzil Sequiera}, \bibinfo{person}{Shashank~S Rao},
  {and} \bibinfo{person}{H~R Shiva~Kumar}.} \bibinfo{year}{2015}\natexlab{}.
\newblock \showarticletitle{Transliteration of {Indic} languages to {Kannada}
  with a user-friendly interface}. In \bibinfo{booktitle}{\emph{Advance
  Computing Conference (IACC), 2015 IEEE International}}. IEEE,
  \bibinfo{pages}{998--1001}.
\newblock


\bibitem[\protect\citeauthoryear{Schultz and Waibel}{Schultz and
  Waibel}{2001}]%
        {schultz2001language}
\bibfield{author}{\bibinfo{person}{Tanja Schultz} {and} \bibinfo{person}{Alex
  Waibel}.} \bibinfo{year}{2001}\natexlab{}.
\newblock \showarticletitle{Language-independent and language-adaptive acoustic
  modeling for speech recognition}.
\newblock \bibinfo{journal}{\emph{Speech Communication}} \bibinfo{volume}{35},
  \bibinfo{number}{1-2} (\bibinfo{year}{2001}), \bibinfo{pages}{31--51}.
\newblock


\bibitem[\protect\citeauthoryear{Van~Heerden, Kleynhans, Barnard, and
  Davel}{Van~Heerden et~al\mbox{.}}{2010}]%
        {datapooling}
\bibfield{author}{\bibinfo{person}{C Van~Heerden}, \bibinfo{person}{N
  Kleynhans}, \bibinfo{person}{E Barnard}, {and} \bibinfo{person}{M Davel}.}
  \bibinfo{year}{2010}\natexlab{}.
\newblock \showarticletitle{Pooling {ASR} data for closely related languages}.
  In \bibinfo{booktitle}{\emph{SLTU 2010: Proc. 2nd Workshop on Spoken
  Languages Technologies for Under-resourced languages}}.
  \bibinfo{pages}{17--23}.
\newblock


\bibitem[\protect\citeauthoryear{Vijay~Girish, Veena, and
  Ramakrishnan}{Vijay~Girish et~al\mbox{.}}{2016}]%
        {vijay2016indicon}
\bibfield{author}{\bibinfo{person}{K~V Vijay~Girish}, \bibinfo{person}{Vijai
  Veena}, {and} \bibinfo{person}{A~G Ramakrishnan}.}
  \bibinfo{year}{2016}\natexlab{}.
\newblock \showarticletitle{Relationship between spoken {Indian} languages by
  clustering of long distance bigram features of speech}. In
  \bibinfo{booktitle}{\emph{India Conference (INDICON), 2016 IEEE Annual}}.
  IEEE, \bibinfo{pages}{1--6}.
\newblock


\end{thebibliography}

\end{document}